\documentclass[12pt]{iopart}
\usepackage[utf8]{inputenc}
\usepackage{graphicx}
\usepackage{color}
\usepackage{hyperref}
\usepackage{bm}
\expandafter\let\csname equation*\endcsname\relax
\expandafter\let\csname endequation*\endcsname\relax
\usepackage{amsmath}

\usepackage{amsmath,amssymb}
\usepackage{ulem}
\usepackage{caption}
\usepackage{cite}
\newcommand{\diff}{\mathrm{d}}
\newcommand{\p}{\partial}

\newcommand{\ele}{\mathrm{e}}
\newcommand{\ion}{\mathrm{i}}

\newcommand{\vthe}{v_{\mathrm{th,e}}}

\newcommand{\para}{\parallel}

\newcommand{\dv}{\diff\bm{v}}
\newcommand{\dr}{\diff\bm{r}}

\newcommand{\Rblacket}[2]{\langle{#1}\rangle_{\bm{R}_{#2}}}

\renewcommand{\footnoterule}{%
  \kern -3pt
  \hrule width \textwidth
  \kern 2.6pt
}

\begin{document}
\title%[]
{
Self-organization of magnetic islands and zonal magnetic fields driven by slab microtearing modes in the absence of magnetic shear
}
\markboth{Self-organization of magnetic islands and zonal magnetic fields}{}

%\author{Mitsuyoshi Yagyu$^1$ \& Ryusuke Numata$^2$}
\author{Mitsuyoshi Yagyu$^1$, Ryusuke Numata$^2$ \& Seiya Nishimura$^3$}
\address{$^1$Faculty of Information and Data Science, Department of Information and Data Science, Notre Dame Seishin University, 2-16-9, Ifukucho, Kita-ku, Okayama-city, Okayama 700-8516, Japan}
\address{$^2$Graduate School of Information Science, University of Hyogo, 
7-1-28 Minatojima Minami-machi, Chuo-ku, Kobe, Hyogo 650-0047, Japan}
\address{$^3$National Institutes for Quantum Science and Technology, Naka 311-0193, Japan}
\ead{mitsu.yagyu@gmail.com}
\vspace{10pt}
%\begin{indented}
%\item[] 2021
%\end{indented}

% https://ja.overleaf.com/project/62590b500dc2d7f94a6bab6f
\begin{abstract}
    Self-organization in collisional microtearing turbulence is investigated using gyrokinetic simulations in a slab geometry without equilibrium magnetic shear. Unlike magnetically driven conventional tearing instabilities, microtearing instabilities are driven by the electron temperature gradient. 
    Electron temperature fluctuations are found to contain most of the gyrokinetic free energy. 
    Although magnetic perturbations contribute little to the free energy, the magnetic field structure plays a crucial role in the nonlinear evolution of microtearing turbulence. 
    Zonal magnetic fields emerge spontaneously through the nonlinear interaction between magnetic perturbations and electron temperature fluctuations. These zonal fields provide an effective magnetic shear that enables magnetic islands to form despite the absence of equilibrium magnetic shear. 
    The electron temperature profile is flattened within the islands. 
    This flattening reduces the instability drive and is expected to promote turbulence saturation.
\end{abstract}

% \begin{abstract}
% The self-organization mechanism in collisional microtearing turbulence driven by an electron temperature gradient is studied using gyrokinetic simulations in the absence of magnetic shear.
% \modify{The energy associated with isotropic electron temperature fluctuations is dominant in microtearing turbulence, while that associated with the temperature anisotropy is negligible.}
% Quantitative analyses demonstrate that zonal magnetic fields are self-generated by the nonlinear interaction between magnetic perturbations and electron temperature fluctuations \modify{and act as an effective magnetic shear, leading to the formation of magnetic islands even in the absence of equilibrium magnetic shear.}
% \modify{It is also shown that the electron temperature profile is flattened within the self-generated magnetic islands. This flattening indicates a reduction in the linear drive of the microtearing modes and may contribute to turbulence saturation, as in microtearing turbulence with finite magnetic shear.}
%\end{abstract}
%

\vspace{2pc}
\noindent{\it Keywords}: electromagnetic turbulence, microtearing modes, zonal magnetic field, self-organization
%\maketitle

\section{Introduction}
\label{sec:introduction}

Microturbulence driven by electron pressure gradients causes anomalous electron heat transport in magnetic confinement devices.
Understanding the heat transport is a crucial issue in the design of future fusion reactors. 
In conventional low-$\beta$ devices ($\beta$ is the ratio of the plasma pressure to the magnetic pressure), electrostatic turbulence driven by the electron temperature gradient (ETG) mode and the trapped electron mode (TEM) is known to contribute to electron heat transport (see, e.g.,~\cite{Dorland+2000,Dannert+2005}).
In higher-$\beta$ devices, electromagnetic instabilities such as microtearing and kinetic ballooning modes can also cause anomalous electron heat transport through magnetic fluctuations.

Numerous gyrokinetic simulations have shown that microtearing modes become unstable and contribute to electron heat transport in existing devices, such as spherical tokamaks (STs)~\cite{Applegate+2004,Wong+2007, Guttenfelder+2012PoP,Dickinson+2013, Patel+2025}, conventional tokamaks~\cite{Doerk+2012, Jian+2019, Hatch+2021}, reversed-field pinches (RFPs)~\cite{Predebon+2013, Carmody+2013}, and stellarators~\cite{Cu+2026, Merlo+2026}, as well as in a conceptual ST~\cite{Patel+2022, Kennedy+2023, Giacomin+2024}.

The microtearing mode changes the topology of the magnetic field and generates small-scale magnetic islands. The islands may overlap to provide a stochastic magnetic field structure, thereby enhancing the electron heat transport along the field lines~\cite{R&R1978, Doerk+2011, Giacomin+2023}.
Nonlinear gyrokinetic simulations using experimental parameters of the National Spherical Torus Experiment (NSTX) showed that electron heat transport driven by microtearing turbulence increases with electron collisionality~\cite{Guttenfelder+2012PoP}, consistent with the observed NSTX confinement scaling~\cite{Kaye+2007}.

The microtearing mode is an ion-scale tearing instability and was first proposed theoretically by Hazeltine et al.~\cite{Hazeltine+1975} using a kinetic model with the rigorous Fokker--Planck collision operator.
They revealed that the velocity dependence of the collision frequency is indispensable for the temperature gradient to drive the microtearing instability.
Subsequently, the dependence of the microtearing mode on the electron--ion collision frequency was investigated theoretically~\cite{Drake+1977} and numerically~\cite{Gladd+1980,Dippolito+1980}.
These studies indicated that the mode is stable in both the collisional and collisionless limits.
Based on the analysis of Drake and Lee~\cite{Drake+1977}, the destabilizing effect is due to pitch-angle scattering with a velocity-dependent collision frequency in the presence of electron temperature nonuniformity.
In the fluid description, this destabilizing effect is represented as the time-dependent thermal force~\cite{Hassam1980_1}.
Through this force, the electron temperature nonuniformity generates a parallel inductive electric field that lags in phase behind the perturbed magnetic field~\cite{Yagyu+2023}.
This phase relation is analogous to the relation between density fluctuations and the electrostatic potential in electrostatic drift waves.
The microtearing mode is fundamentally an electromagnetic drift wave and is destabilized solely by the time-dependent thermal force even in the absence of equilibrium magnetic shear~\cite{Hassam1980_2}.
%In this shearless case, the instability is referred to as the self-filamentation mode~\cite{Hassam1980_2}.
Electromagnetic microinstabilities driven by the electron temperature gradient have been reported even in the collisionless regime, both in slab geometry and in toroidal geometry.
These modes have been discussed as a tearing-parity electromagnetic ETG mode~\cite{Zocco+2015,Geng+2020} and a collisionless microtearing mode~\cite{Catto+1981,Applegate+2007,Chandran+2024}, respectively.
Such collisionless modes are outside the scope of this paper.

The nonlinear saturation of microtearing turbulence has been investigated since the early drift-kinetic analysis of a slab plasma without magnetic shear by Drake et al.~\cite{Drake+1980}.
They showed that electron streaming along the perturbed magnetic field $\bm{B}_\perp$ transfers energy from shorter- to longer-wavelength modes.
Saturation occurs when the stabilizing effect of this nonlinear energy transfer balances the linear growth.
The resulting saturation amplitude scales as $|\bm{B}_\perp|/B_0 \sim \rho_{\mathrm{e}}/L_{T_{0\mathrm{e}}}$, where $\rho_{\mathrm{e}}$ and $L_{T_{0\mathrm{e}}}$ are the electron Larmor radius and temperature gradient scale length, respectively.
This scaling was reproduced by nonlinear gyrokinetic simulations~\cite{Doerk+2011}.
Several numerical studies have reported different saturation mechanisms for microtearing turbulence. 
One such mechanism is the local flattening of the electron temperature profile at low-order rational surfaces, which has been shown to reduce the linear drive and to play an important role in determining the saturation level of the electron heat flux~\cite{Ajay+2023}.

The generation of zonal magnetic fields has been proposed as another saturation mechanism for microtearing turbulence.
%Zonal magnetic fields are another candidate for the saturation mechanism.
Some gyrokinetic simulations of microtearing turbulence have demonstrated that zonal magnetic fields play an important role in nonlinear saturation, as removing the zonal magnetic field component leads to a significant increase in the electron heat flux~\cite{Pueschel+2020,Giacomin+2023}. In contrast, other studies using gyrokinetic simulations have reported that zonal fields do not play a significant role in nonlinear saturation~\cite{Doerk+2011,Ajay+2023,Hamed+2025}.
The role of zonal magnetic fields in microtearing turbulence remains controversial.
We note that zonal magnetic fields were not explicitly considered in the saturation theory of Drake et al.~\cite{Drake+1980}.

We investigate the self-organization mechanism of zonal magnetic fields in microtearing turbulence using the configuration without magnetic shear considered by Drake et al.~\cite{Drake+1980}.
This configuration retains the destabilization mechanism of the collisional microtearing mode, i.e., the time-dependent thermal force, since magnetic shear acts only to stabilize the mode~\cite{Drake+1977}.
A region of weak or nearly zero magnetic shear is realized in magnetic confinement devices as well, e.g., in stellarators~\cite{Faber+2018,Cu+2026, Mulholland+2026} and in advanced tokamaks~\cite{Jian+2021, Wakatsuki+2023,Gabriellini+2025}.
The insights obtained in this study are expected to contribute to the understanding of electromagnetic turbulence in weak- or nearly-zero-magnetic-shear regimes.

In this paper, we present numerical analyses of collisional microtearing turbulence in the absence of magnetic shear using a gyrokinetic model. 
We restrict our analyses to slab geometry as the simplest problem setup, so that toroidal effects such as the magnetic drift and trapped particle effects are excluded. 
We show that zonal magnetic fields are self-generated by the nonlinear interaction between magnetic perturbations and electron temperature fluctuations. The zonal magnetic fields act as an effective magnetic shear, leading to the formation of magnetic islands even in the absence of equilibrium magnetic shear.
We also show that the electron temperature profile is flattened within the self-generated magnetic islands.
Regardless of the presence of magnetic shear, this flattening weakens the local temperature-gradient drive of the microtearing instability and is expected to promote turbulence saturation.

The remainder of this paper is organized as follows.
In section~\ref{sec:gyrokinetic_linear_simulation}, we describe the gyrokinetic model and the simulation setup, and show that the microtearing modes are unstable in the absence of magnetic shear.
In section~\ref{sec:nonlinear_evolution}, we present the nonlinear simulation results, including the analysis of the gyrokinetic free energy in terms of fluid moments, the formation of magnetic islands in the absence of magnetic shear, and the generation mechanism of zonal magnetic fields. The flattening of the electron temperature profile within the magnetic islands is also shown in the same section.
Finally, we summarize the paper in section~\ref{sec:summary}.
%The gyrokinetic--Maxwell equations and the derivation of the power balance equation for $A_{\para}$ are given in \ref{appdix_gk_framework} and \ref{appdix_ohms_law}, respectively.
\section{Gyrokinetic framework and linear instability}
\label{sec:gyrokinetic_linear_simulation}

\subsection{Model and simulation setup}
\label{subsec:model_simulationsetup}

We solve the electromagnetic gyrokinetic equations of ions and electrons for the non-Boltzmann part $h_{s}$ of the perturbed distribution function $\delta f_{s}$, where $\delta f_{s} = -(q_{s}\phi/T_{0s}) f_{0s} + h_{s}$, and the subscript $s$ is the species label ($s = \mathrm{i}, \mathrm{e}$).
The gyrokinetic equation and the definitions of the associated symbols are given in \ref{appdix_GK-M_eqn}.
The finite Larmor radius (FLR) effects are fully retained for ions. Electron FLR effects are also retained, but they are small for the parameters specified below.
The parallel magnetic perturbation $\delta B_\parallel$ is not included in this paper.

We perform gyrokinetic simulations of electromagnetic turbulence driven by collisional microtearing modes in a 2--D slab geometry using the \texttt{GS2} code~\cite{GS2_8_2_1}.
We consider a non-uniform plasma with electron pressure $P_{0\ele}(x)$ embedded in an equilibrium magnetic field $\bm{B}_{0}=B_{0} \hat{\bm{z}} + B_{y}^{\mathrm{eq}}(x)\hat{\bm{y}}$.
Here, $B_{y}^{\mathrm{eq}}(x) = B_{0}(x/L_{s})$ is the sheared field component with $L_{s}$ being the magnetic shear scale length.
%Throughout this paper, magnetic shear refers to that of the equilibrium magnetic field. 
Unless otherwise stated, we consider the configuration without magnetic shear, i.e., $L_{s}^{-1} = 0$.
The electron pressure gradient is taken in the $x$ direction and is characterized by $L_{P_{0\ele}}^{-1} = - \p_{x} \ln P_{0\ele}(x)$. The corresponding diamagnetic drift frequency is given by $\omega_{\ast,\ele} = \omega_{\ast,n_{0\ele}} + \omega_{\ast,T_{0\ele}}$, where $\omega_{\ast,n_{0\ele}}$ and $\omega_{\ast,T_{0\ele}}$ are the diamagnetic drift frequencies associated with the density gradient $L_{n_{0\ele}}^{-1}$ and temperature gradient $L_{T_{0\ele}}^{-1}$, respectively. The definitions of $\omega_{\ast,n_{0\ele}}$ and $\omega_{\ast,T_{0\ele}}$ are given in \ref{appdix_GK-M_eqn}.
The characteristic time scale of the system is determined by $\omega_{\ast,\ele}^{-1}$. In this paper, $\omega_{\ast,\ele}$ is evaluated at $k_y\rho_{\ion}=1$.
With the parameters presented later, $\omega_{\ast,\ele}^{-1}$ is related to the commonly used time scale $L_{T_{0\ele}}/v_{\mathrm{th},\ion}$ by $\omega_{\ast,\ele} (L_{T_{0\ele}}/v_{\mathrm{th},\ion}) = 0.6$.
The time variable is normalized by $\omega_{\ast,\ele}^{-1}$ and denoted by $\hat{t}$ throughout this paper.

The physical parameters are set as follows: the electron plasma beta $\beta_{\ele}\equiv n_{0\ele}T_{0\ele}/(B_{0}^{2}/2\mu_{0})=0.01$, the temperature ratio $T_{0\ion}/T_{0\ele}=1$, the density ratio $n_{0\ion}/n_{0\ele}=1$, the mass ratio $m_{\ele}/m_{\ion} = 1/1836$, the normalized temperature gradients $L_{n_{0\ele}}/L_{T_{0\ele}}=5$ and $L_{n_{0\ion}}/L_{T_{0\ion}} =0$, and the ion charge number $-q_{\ion}/q_{\ele}=1$.
\texttt{GS2} employs the model Fokker--Planck collision operator~\cite{Abel+2008,Barnes+2009}, which includes pitch-angle scattering, energy diffusion, and terms that ensure the conservation of the particle number, momentum, and energy.
%Collisions are \modify{described by} the model Fokker--Planck collision operator, which includes pitch-angle scattering, energy diffusion, and terms that ensure the conservation of the particle number, momentum, and energy.
The electron collisionality is set to $\nu_{\ele\ele}/\omega_{\ast,\ele} = 0.15$.
Since we assume a single ion species with $-q_{\ion}/q_{\ele} = 1$, we have $\nu_{\mathrm{ee}} = \nu_{\mathrm{ei}}$.
We set the ion collision frequency to be much smaller than that for electrons in this paper.

\subsection{Microtearing instability in the absence of magnetic shear}
\label{subsec:mt_linear_results}

%We perform linear simulations to identify the unstable regime of microtearing modes in $k_{y}\rho_\ion$ space.
We perform linear simulations to identify the range of $k_y\rho_{\ion}$ over which microtearing modes are unstable.
In linear runs, we set the numerical resolution in the velocity space to $(N_{\lambda}, N_{E}) = (8, 16)$, where $N_{\lambda}$ and $N_{E}$ denote the numbers of grid points in the pitch-angle and energy coordinates, respectively~\cite{Barnes+2010}.

\begin{figure}[htp]
    \centering
    \includegraphics[width=1.0\linewidth]{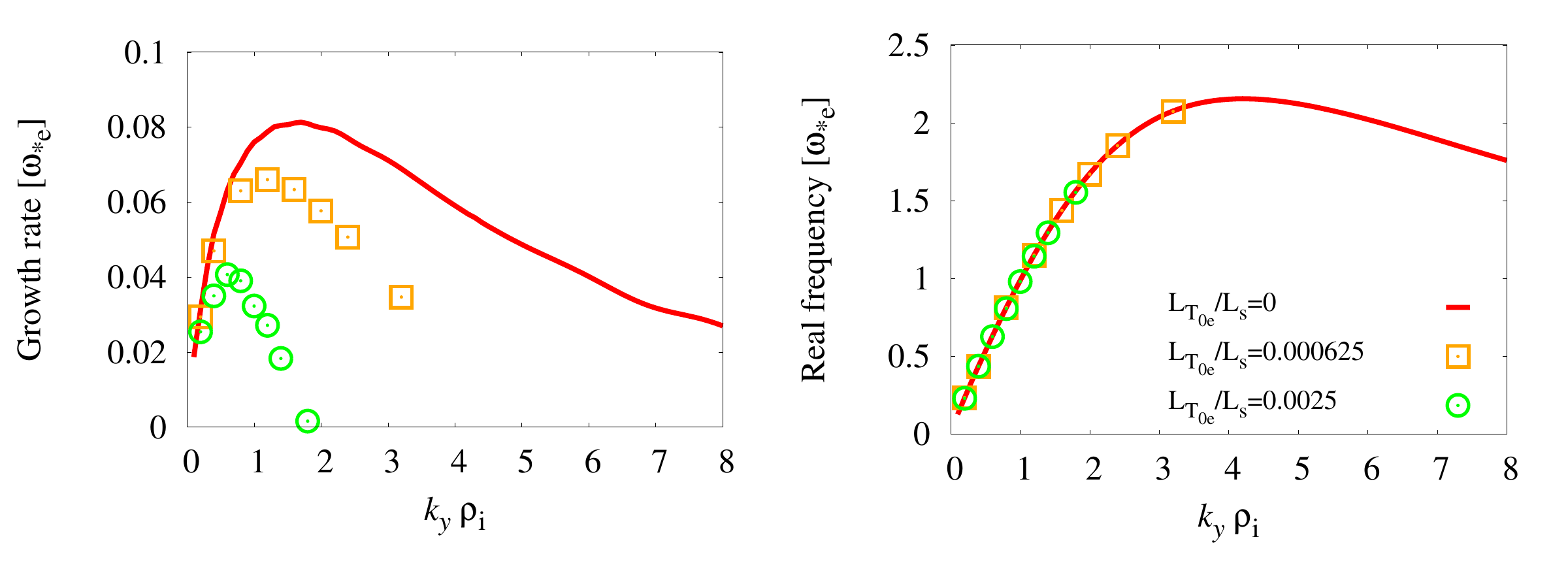}
    \caption{Growth rate (left) and real frequency (right) of microtearing modes as a function of $k_{y} \rho_{\ion}$.}
    \label{fig:eigenvalues}
\end{figure}

Figure~\ref{fig:eigenvalues} shows the linear growth rate (left) and real frequency (right) of microtearing modes as a function of $k_y \rho_{\ion}$. The eigenvalues are normalized by $\omega_{\ast,\ele}$.
The solid red lines denote the eigenvalues without magnetic shear.
The microtearing modes become unstable even in the absence of magnetic shear (left) and propagate in the direction of the electron diamagnetic drift (right).
For comparison, the eigenvalues with finite magnetic shear are also shown by the green circles and orange squares.
The magnetic shear becomes stronger as $L_{T_{0\ele}}/L_{s}$ increases at fixed $L_{T_{0\ele}}$.
The real frequency is unaffected by magnetic shear, while the growth rate is suppressed by its stabilizing effect, consistent with theoretical~\cite{Drake+1977} and numerical~\cite{Jian+2021, Cu+2026} studies.
In the finite magnetic shear case, sufficient grid points along the equilibrium magnetic field are used to resolve the fine structure of the mode. We have confirmed the convergence of the results.

\section{Nonlinear evolution and self-organization in microtearing turbulence}
%\subsection{Self-organization mechanism of magnetic islands and zonal magnetic fields}
\label{sec:nonlinear_evolution}

In this section, we present the results of nonlinear simulations in the absence of magnetic shear.
The simulations are performed with the numerical resolution $(N_{x},N_{y}, N_{\lambda},N_{E}) = (64,64,8,16)$ in the domain of $L_{x}/\rho_{\ion}= L_{y}/\rho_{\ion} = 10\pi$.
Artificial hyperviscosity is included to dissipate the energy at small scales and to achieve a steady state.
Its amplitude and cut-off wavenumber are chosen to be $D_{\mathrm{hyp}}/\omega_{\ast,\ele}\approx 0.3$ and $k_{x,y}^{\mathrm{cut}} \rho_{\ion} \approx 0.9\, k_{x,y}^{\mathrm{max}} \rho_{\ion}$, respectively, where $k_{x,y}^{\mathrm{max}} \rho_{\ion} \approx 2\pi (\rho_{\ion}/L_{x,y})(N_{x,y}/3)$ is the maximum wavenumber.
As shown in section~\ref{subsec:energy_analysis}, reaching nonlinear saturation requires long simulation times, which makes high-resolution simulations computationally expensive.
We therefore use a lower spatial resolution in this paper.
We have confirmed that the conclusions of this paper do not change with resolution and hyperviscosity.

% \modify{
% In section~\ref{subsec:energy_analysis}, we show the time evolution of the gyrokinetic free energy and identify the dominant fluid-moment contribution to the free energy.
% We demonstrate the formation of magnetic islands by zonal magnetic fields in the absence of magnetic shear (section~\ref{subsec:magnetic_island}) and elucidate the generation mechanism of the zonal magnetic fields (section~\ref{subsec:zonal_field}).
% Finally, in section~\ref{subsec:temperature_flattening}, we show the flattening of the electron temperature profile within the self-generated magnetic islands.
% }

\subsection{Analysis of energy evolution and fluid-moment contributions}
\label{subsec:energy_analysis}
We first consider the energetics of the system. The gyrokinetic free energy $W$ consists of the particle part $K_{s}$ and the magnetic field part $M_{\perp}$, and is given by
\begin{align}
	W = \sum_{s} K_{s} + M_{\perp}
	= \frac{1}{L_{x} L_{y}}\int \left[ 
	\sum_{s} \int \frac{T_{0s} |\delta f_{s}|^{2}}{2 f_{0s}}\dv
	+ \frac{|\nabla_{\perp} A_{\para}|^2}{2\mu_{0}}
	\right] \diff \bm{r}.
    \label{eq:generalized_energy}
\end{align}
The subscripts $\para$ and $\perp$ denote the components parallel and perpendicular to the equilibrium magnetic field, respectively.
We decompose the particle energy $K_{s}$ into contributions from low-order fluid moments by expanding $\delta f_{s}$ in Hermite--Laguerre polynomials
\begin{align}
    K_{s} \approx
    K^{\tilde{n}}_{s} + K^{\tilde{u}_{\para}}_{s}
    + K^{\tilde{T}}_{s} + K^{\tilde{T}_{\para,\perp}}_{s},
    \label{eq:fluid_moments_expansion}
\end{align}
where the terms on the right-hand side are defined as follows:
\begin{align}
    &K^{\tilde{n}}_{s} = \frac{1}{L_{x} L_{y}} \int \frac{1}{2} n_{0s}T_{0s}\left| \frac{\tilde{n}_{s}}{n_{0s}} \right|^2 \dr,
    \label{eq:fluid_moments_density}\\
    &K^{\tilde{u}_{\para}}_{s} = \frac{1}{L_{x} L_{y}} \int \frac{n_{0s} m_{s}}{2}|\tilde{u}_{\para,s}|^{2} \dr,\\
    &K^{\tilde{T}}_{s} = \frac{1}{L_{x} L_{y}} \int \frac{3}{4}n_{0s}T_{0s}\left| \frac{\tilde{T}_{s}}{T_{0s}} \right|^2 \dr, \\ 
    &K^{\tilde{T}_{\para,\perp}}_{s} = \frac{1}{L_{x} L_{y}} \int \frac{n_{0s}T_{0s}}{6}\frac{|\tilde{T}_{\para,s}-\tilde{T}_{\perp,s} |^{2}}{T_{0s}^{2}} \dr.
    \label{eq:fluid_moments}
\end{align}
The quantities $\tilde{n}_{s}$, $\tilde{u}_{\para,s}$, $\tilde{T}_{s}$, $\tilde{T}_{\para,s}$, and $\tilde{T}_{\perp,s}$ are the velocity moments of $\delta f_{s}$ defined in \ref{appdix:fluidmoms}.
For simplicity, \eqref{eq:fluid_moments_expansion}--\eqref{eq:fluid_moments} are derived under the drift-kinetic approximation, $k_{\perp}\rho_{s} \to 0$.
%whereas the finite Larmor radius effects are retained in the \texttt{GS2} simulations.

\begin{figure}[htp]
    \centering
    \includegraphics[width=0.9\linewidth]{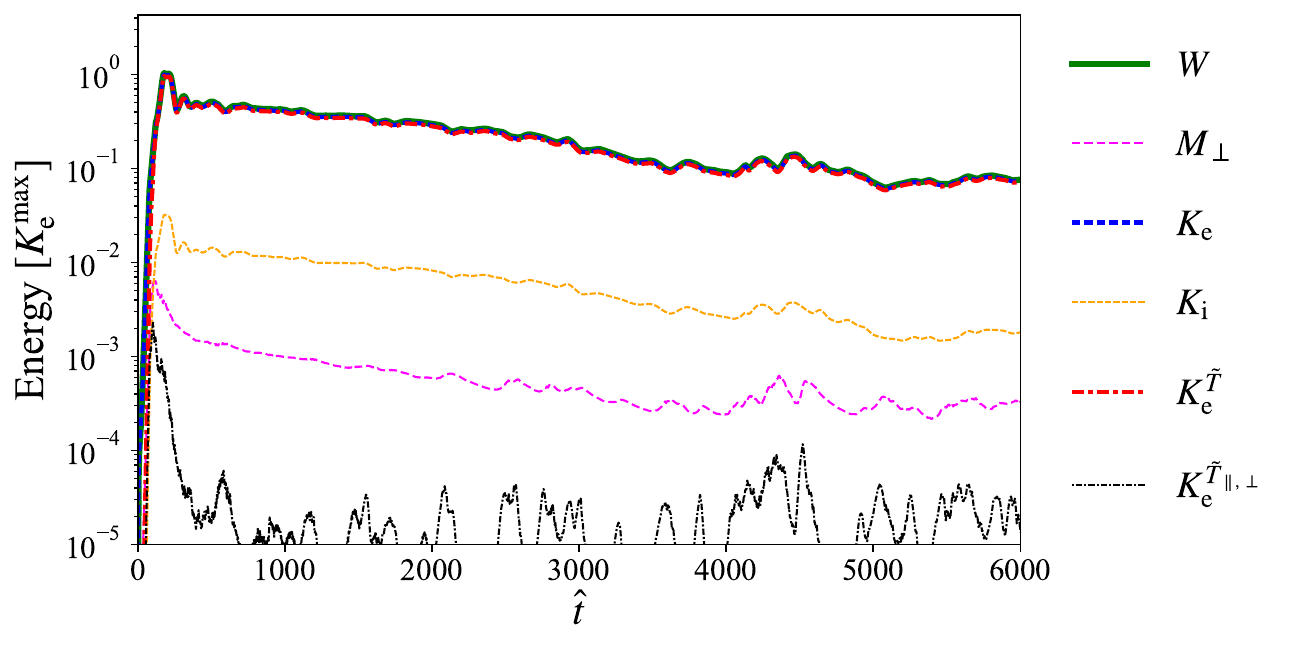}
    \caption{Time evolution of the gyrokinetic free energy and its components. 
    }
    \label{fig:total_energy_evolution}
\end{figure}

Figure~\ref{fig:total_energy_evolution} shows the time evolution of the energies.
All energies are normalized by the maximum value of the electron particle energy over time, $K_{\ele}^{\mathrm{max}} = \max K_{\ele}(\hat{t})$.
Unlike conventional tearing instabilities that release energy stored in the equilibrium magnetic field, microtearing instabilities are driven by the electron temperature gradient. As shown in figure~\ref{fig:total_energy_evolution}, the electron particle energy $K_\ele$ dominates the gyrokinetic free energy, and the magnetic energy $M_\perp$ remains much smaller. Nevertheless, the magnetic field structure plays a crucial role in the nonlinear evolution of the turbulence, as discussed later. Most of $K_\ele$ is associated with isotropic electron temperature fluctuations (see $K_{\ele}^{\tilde{T}}$). 
The energy associated with the temperature anisotropy, $K^{\tilde{T}_{\para,\perp}}_{\ele}$, is much smaller than $K^{\tilde{T}}_{\ele}$ because pitch-angle scattering relaxes the difference between $\tilde{T}_{\para,\ele}$ and $\tilde{T}_{\perp,\ele}$.
Although not shown in the figure, $K^{\tilde{n}}_{\ele}$ and $K^{\tilde{u}_{\para}}_{\ele}$ are also much smaller than $K^{\tilde{T}}_{\ele}$.
%The dominance of $K^{\tilde{T}}_{\ele}$ is consistent with the destabilization mechanism of the microtearing mode.
%The time-dependent thermal force enables the mode to tap the free energy of the equilibrium electron temperature gradient (see section~\ref{sec:introduction}).
%The free energy is thus stored predominantly in the electron temperature fluctuations.

We divide the time evolution of the energy shown in figure~\ref{fig:total_energy_evolution} into four phases.
In the linear phase ($\hat{t} \lesssim 50$), the energy grows exponentially, as it is injected from the background temperature gradient through the microtearing instability.
In the subsequent early nonlinear phase ($50 < \hat{t} \lesssim 600$), the energy reaches its peak as the system transitions from linear growth to nonlinear dynamics.
This phase consists of two stages.
In the zonal-field generation stage ($50 < \hat{t} \lesssim 200$), coherent zonal magnetic fields, uniform in the $y$ direction, are self-generated through the nonlinear interaction between magnetic perturbations and electron temperature fluctuations. In the island formation stage ($200 < \hat{t} \lesssim 600$), these zonal magnetic fields act as an effective magnetic shear, thereby leading to the formation of magnetic islands.
In the intermediate nonlinear phase ($600 < \hat{t} \lesssim 4200$), the energy gradually decays.
Finally, the energy becomes nearly constant in the saturation phase ($\hat{t} > 4200$).
This paper focuses on the early nonlinear phase.
%A detailed analysis of the intermediate nonlinear and saturation phases is left for future work.

\subsection{Self-organization of magnetic islands and zonal magnetic fields}
\label{subsec:magnetic_island}
We next examine the generation of zonal magnetic fields and their role in the formation of magnetic islands.

\begin{figure}[!ht]
    \centering
    \includegraphics[width=0.8\linewidth]{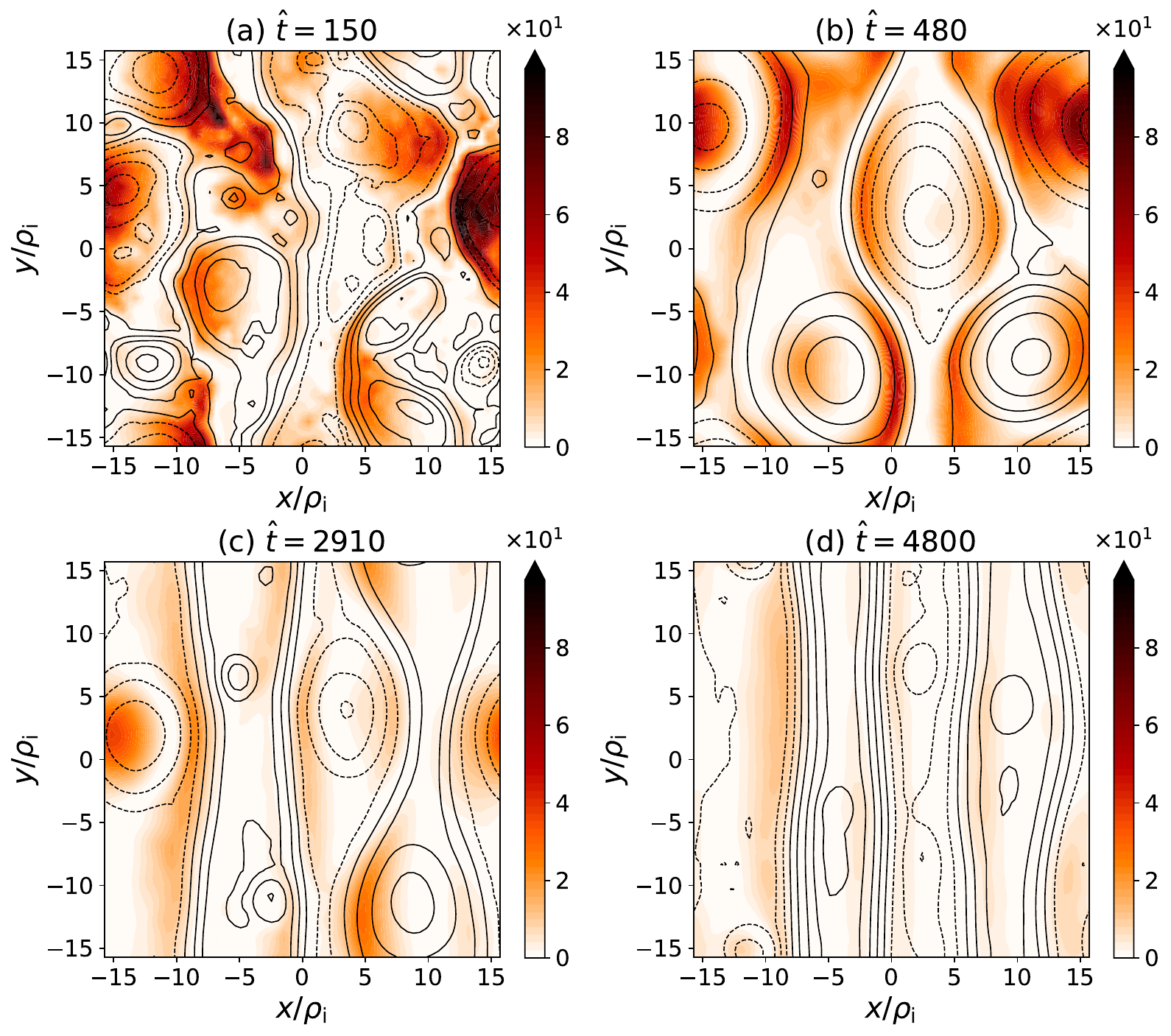}
    \caption{Snapshots of the spatial distribution of the electron temperature fluctuation power $\tilde{T}_{\ele}^{2}$. The contour lines indicate $A_{\para}$.
    }
    \label{fig:temp_amp2d}
\end{figure}

Figure~\ref{fig:temp_amp2d} shows snapshots of the spatial distribution of the power of the electron temperature fluctuation, $\tilde{T}_{\ele}^{2}$, in the $x$--$y$ plane at $\hat{t} = 150$, $480$, $2910$, and $4800$.
$\tilde{T}_{\ele}$ is normalized by $T_{0\ele} \rho_{\ion}/L_{T_{0\ele}}$.
The contour lines of $A_{\para}$, also shown in figure~\ref{fig:temp_amp2d}, represent the perturbed magnetic field lines in the $x$--$y$ plane.
Figure~\ref{fig:temp_amp2d}(a) shows that the magnetic fields are in a turbulent state.
Subsequently, coherent chains of magnetic islands aligned in the $y$ direction are formed (figure~\ref{fig:temp_amp2d}(b)).
Note that these islands are formed even in the absence of magnetic shear.
The island width then gradually decreases (figure~\ref{fig:temp_amp2d}(c)), and the zonal field structure becomes dominant (figure~\ref{fig:temp_amp2d}(d)).
Throughout these processes, $\tilde{T}_{\ele}^{2}$ is localized along the magnetic field lines and gradually decays with time.
This spatial correspondence indicates a coupling between the magnetic perturbations and the electron temperature fluctuations.

\begin{figure}[!ht]
    \centering
    \includegraphics[width=0.6\linewidth]{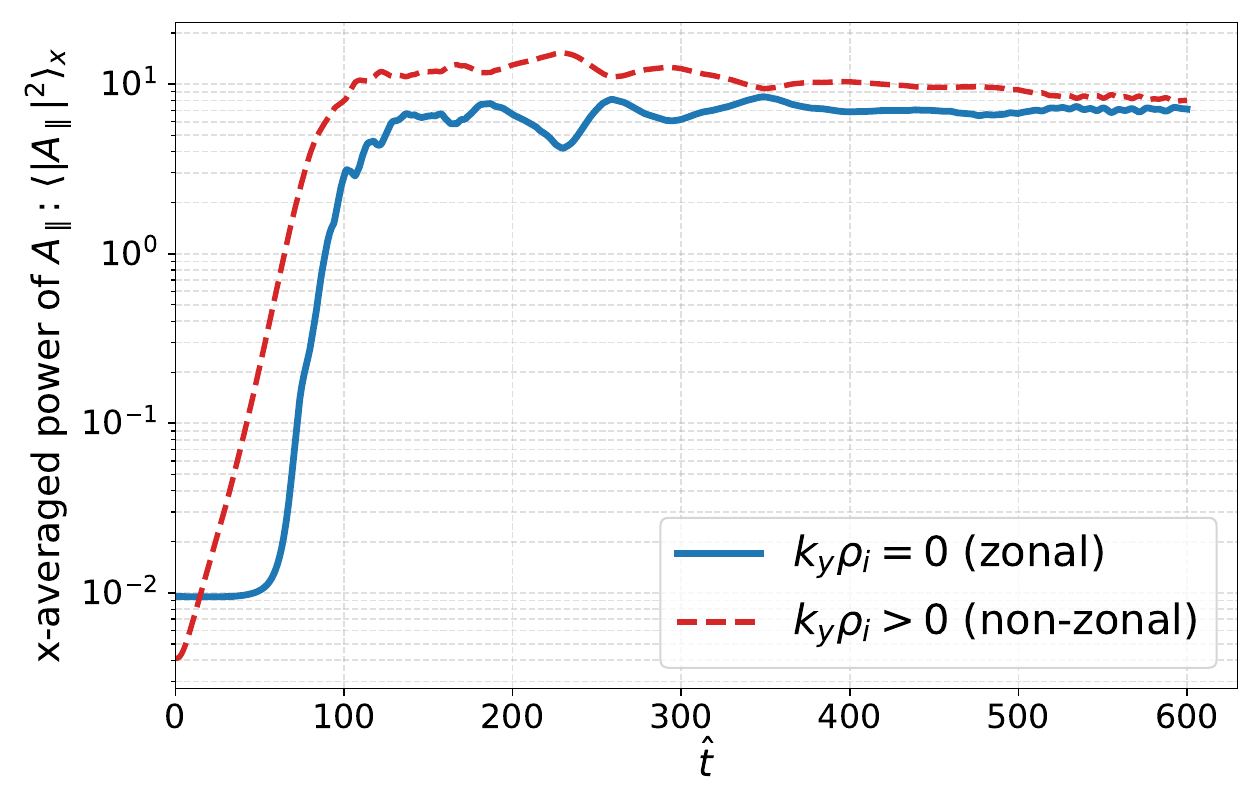}
    \caption{Time evolution of the $x$-averaged power of the zonal and non-zonal components of $A_{\para}$. 
    }
    \label{fig:apar2_ky_tevol}
\end{figure}

Figure~\ref{fig:apar2_ky_tevol} shows the time evolution of the $x$-averaged power of $A_{\para}$ for the zonal component ($k_{y}\rho_{\ion}=0$, solid blue line) and the non-zonal components ($k_{y} > 0$, dashed red line).
$A_{\para}$ is normalized by $(q_{\ele}\vthe/T_{0\ele})^{-1}(\rho_{\ion}/L_{T_{0\ele}})$.
To focus on the island formation, we show the time evolution for $\hat{t} \leq 600$.
The non-zonal components grow exponentially in the linear phase ($\hat{t} \lesssim 50$), and their growth stops around $\hat{t} \approx 100$.
The zonal component grows much more rapidly for $50 < \hat{t} \lesssim 100$ and reaches an amplitude comparable to that of the non-zonal components.

\begin{figure}[!ht]
    \centering
    \includegraphics[width=0.9\linewidth]{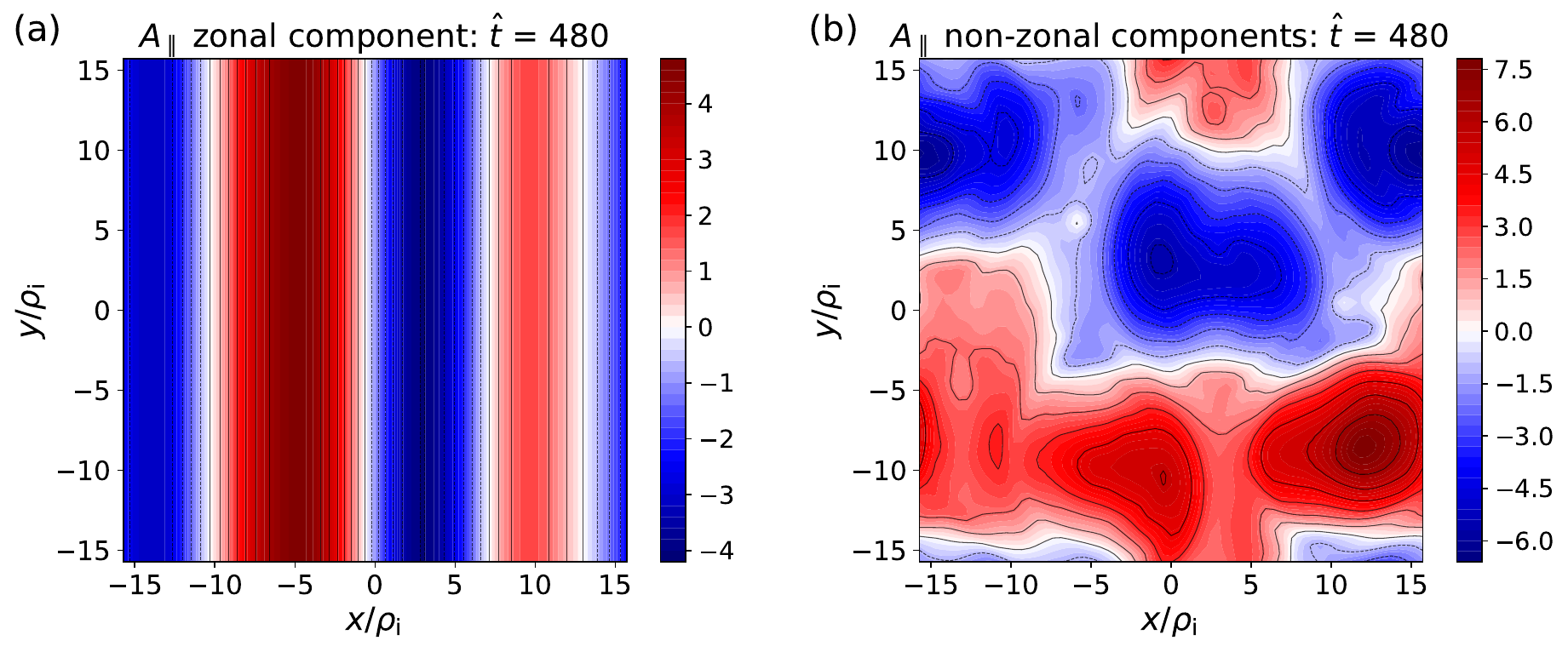}
    \caption{A snapshot of $A_{\para}$ for (a) the zonal component and (b) the non-zonal components at $\hat{t}=480$.
    The contour lines also denote $A_{\para}$.
    }
    \label{fig:zonal-nonzonal_apar}
\end{figure}

\begin{figure}[!ht]
    \centering
    \includegraphics[width=0.6\linewidth]{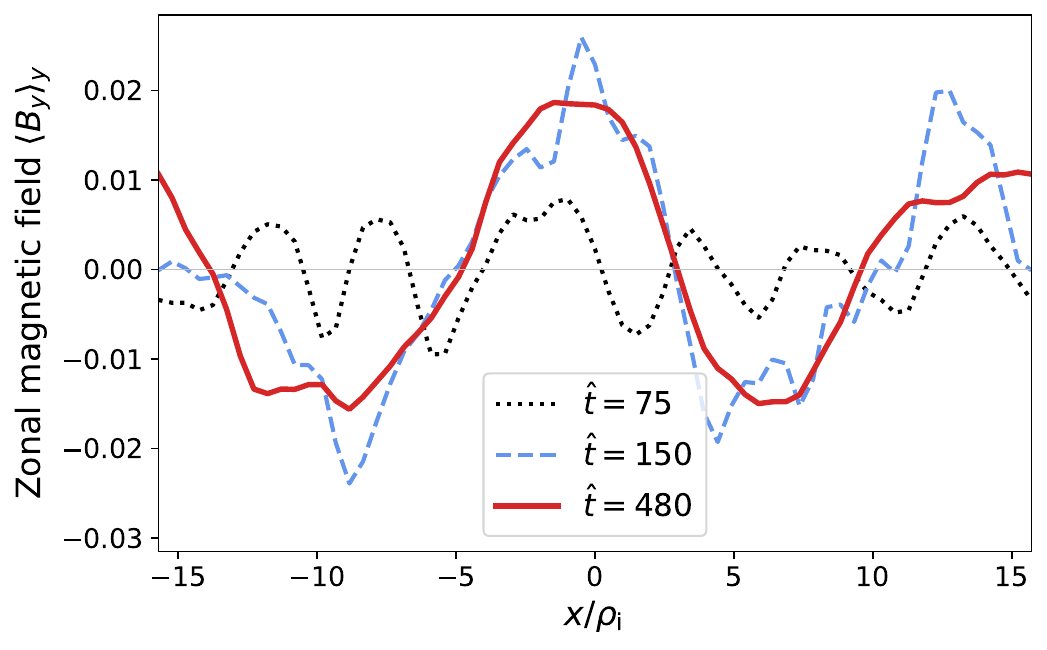}
    \caption{Zonal magnetic field profile in the $x$ direction at different times. $\langle B_{y} \rangle_{y} = -\p \langle A_{\para} \rangle_{y} /\p x$ is plotted, where $\langle \cdot \rangle_{y}$ denotes the $y$-average. 
    }
    \label{fig:by_profile_steady_state}
\end{figure}

We decompose $A_\parallel$ in the $x$--$y$ plane at $\hat{t}=480 $ into its zonal and non-zonal components, shown in figures~\ref{fig:zonal-nonzonal_apar}(a) and \ref{fig:zonal-nonzonal_apar}(b), respectively.
The contour lines also denote $A_{\para}$.
The chains of magnetic islands aligned in the $y$ direction are absent when the zonal component is removed (figure~\ref{fig:zonal-nonzonal_apar}(b)), highlighting the role of zonal magnetic fields in their formation.
Figure~\ref{fig:by_profile_steady_state} shows the $x$ profile of the zonal magnetic field $\langle B_{y} \rangle_{y} = -\p \langle A_{\para} \rangle_{y}/\p x$ at different times, where $\langle \cdot \rangle_{y}$ denotes the $y$-average. $B_{y}$ is normalized by $B_{0} \rho_{\ion}/L_{T_{0\ele}}$.
%\modify{The zonal magnetic fields have a small-scale structure ($\hat{t}=75$), grow rapidly ($\hat{t}=150$), and form a coherent structure ($\hat{t}=480$) in the $x$ direction.}
The O-points of the magnetic islands are located near the $x$ positions where $\langle B_{y} \rangle_{y} = 0$.
The turbulence thus generates its own magnetic shear, allowing the magnetic perturbations to self-organize into island chains even in a shearless equilibrium.
%The self-generated zonal magnetic fields act as an effective magnetic shear, leading to the formation of magnetic islands even in the absence of equilibrium magnetic shear.

To identify the source of the zonal magnetic fields, we analyze the power balance of \(A_\parallel\) using the generalized Ohm's law.
%To elucidate the generation mechanism of the zonal magnetic fields, we analyze the power balance of $A_{\para}$ derived from the generalized Ohm's law.
The growth rate of $|A_{\para}|$ at each $k_y$ is decomposed into contributions from the individual terms in the generalized Ohm's law:
\begin{align}
    \hat{\gamma}_{k_y} = \sum_j \hat{\gamma}^{j}_{k_y},
    \label{eq:growth_rate_decomposition_main}
\end{align}
where the hat on $\gamma$ denotes normalization by $\omega_{\ast,\ele}$. The index $j$ labels the contributions from electron inertia, $\bm{E}\times\bm{B}$ advection, nonlinear interactions between magnetic perturbations and density and temperature fluctuations, and the collision term.
The detailed derivation of \eqref{eq:growth_rate_decomposition_main} is given in \ref{appdix_ohms_law}.

\begin{figure}[!ht]
    \centering
    \includegraphics[width=0.6\linewidth]{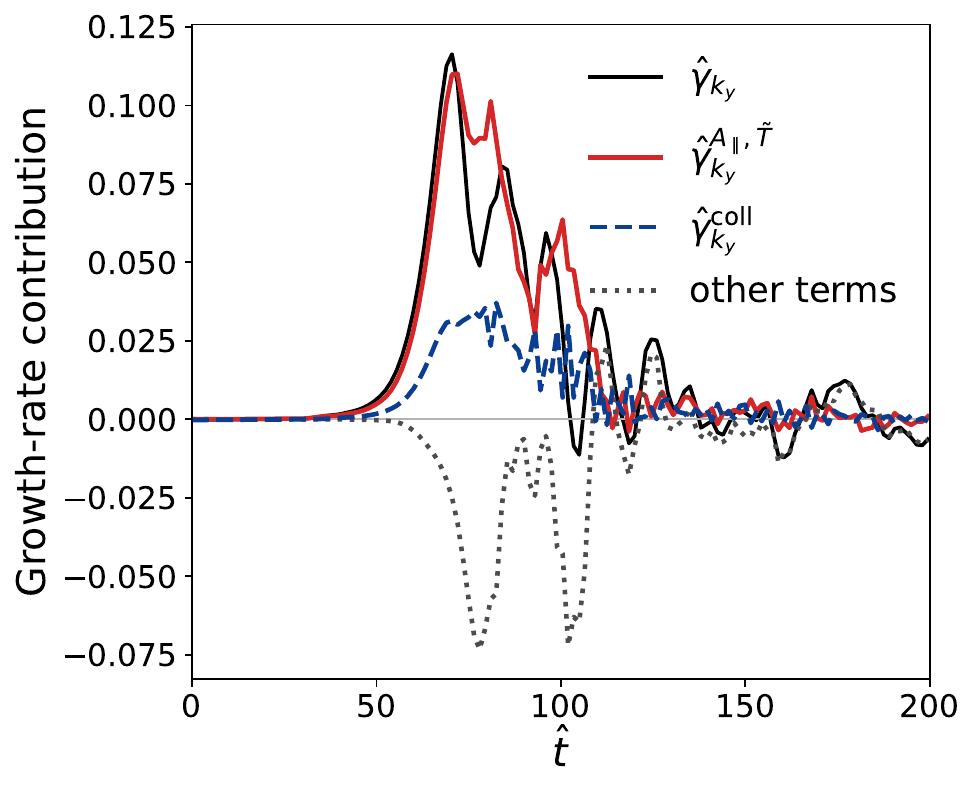}
    \caption{Time evolution of the growth rate of $|A_{\parallel}|$ for the zonal component ($k_{y}\rho_{\ion}=0$).}
    \label{fig:apar2_tevol_ky0}
\end{figure}

Figure~\ref{fig:apar2_tevol_ky0} shows the time evolution of each contribution to the growth rate of the zonal component ($k_y\rho_{\ion}=0$) during the linear phase and the zonal-field generation stage ($\hat{t} \leq 200$).
The solid black, solid red, and dashed blue lines represent $\hat{\gamma}_{k_{y}}$, $\hat{\gamma}^{A_{\para},\tilde{T}}_{k_{y}}$, and $\hat{\gamma}^{\mathrm{coll}}_{k_{y}}$, respectively, and the dotted gray line represents the sum of the other terms.
For $\hat{t} \lesssim 50$, $\hat{\gamma}_{k_y} \approx 0$ because the zonal component has no linear drive. The zonal component grows rapidly for $50 < \hat{t} \lesssim 100$, driven primarily by $\hat{\gamma}^{A_{\para},\tilde{T}}_{k_y}$. The term $\hat{\gamma}^{A_{\para},\tilde{T}}_{k_y}$ represents the nonlinear interaction between magnetic perturbations and electron temperature fluctuations in the generalized Ohm's law. Zonal magnetic fields are generated by this nonlinear interaction.

% The dominance of $\hat{\gamma}^{A_{\para},\tilde{T}}_{k_{y}}$ demonstrates that zonal magnetic fields are self-generated by the turbulence through the nonlinear interaction between magnetic perturbations and electron temperature fluctuations.
% Because the zonal component is driven by the products of the non-zonal modes, it starts to grow abruptly once the non-zonal modes reach sufficiently large amplitudes, and it grows faster than the non-zonal modes (figure~\ref{fig:apar2_ky_tevol}).

\subsection{Electron temperature profile flattening within magnetic islands}
\label{subsec:temperature_flattening}

\begin{figure}[!ht]
    \centering
    \includegraphics[width=0.5\linewidth]{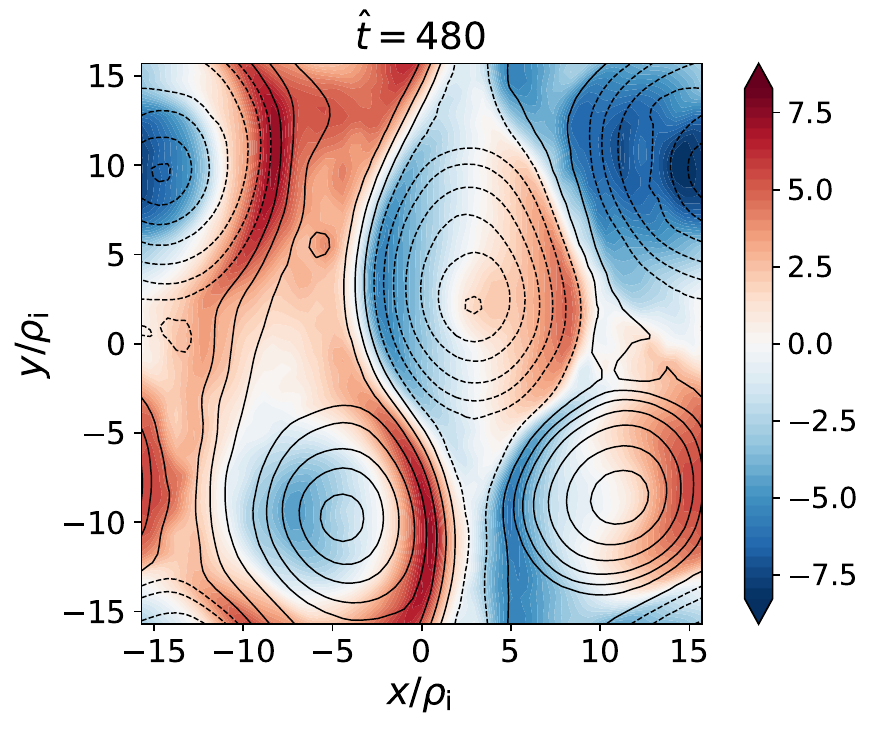}
    \caption{Snapshot of the electron temperature fluctuation during the formation of magnetic islands. The contour lines denote $A_{\para}$.
    }
    \label{fig:temperature_flattening}
\end{figure}

In this section, we analyze the electron temperature fluctuation profile within the self-generated magnetic islands.
Figure~\ref{fig:temperature_flattening} shows the spatial distribution of $\tilde{T}_{\ele}$ at $\hat{t}=480$. The contour lines represent $A_{\para}$.
Remarkably, the temperature fluctuations spontaneously organize within the self-generated magnetic islands to produce a gradient that nearly cancels the background temperature gradient.
%The temperature fluctuations within the magnetic islands tend to form a gradient in the opposite direction to the background temperature gradient.
We focus on the magnetic island centered at $(x/\rho_{\ion}, y/\rho_{\ion}) \approx (2.9, 2.5)$.
The island width is $w \approx 9.8\rho_{\ion}$.
With the normalized temperature fluctuation $\hat{T}_{\ele} = \tilde{T}_{\ele}/(T_{0\ele}\rho_{\ion}/L_{T_{0\ele}})$ and the normalized coordinate $\hat{x} = x/\rho_{\ion}$, $\hat{T}_{\ele}$ varies from approximately $-5$ to $5$ across the island, so that $\p \hat{T}_{\ele}/\p \hat{x} = (L_{T_{0\ele}}/T_{0\ele})\, \p \tilde{T}_{\ele}/\p x \approx 1$.
In the same normalization, the background temperature gradient is $(L_{T_{0\ele}}/T_{0\ele})\,\p T_{0\ele}/\p x = -1$, where $\p T_{0\ele}/\p x = -T_{0\ele}/L_{T_{0\ele}}$.
%\modify{Within the islands, the gradient of the temperature fluctuations cancels the background temperature gradient, flattening the temperature profile.}
%The temperature fluctuations are distributed so as to flatten the electron temperature profile within the magnetic island.
Although the above estimate is based on a single island in the snapshot shown in figure~\ref{fig:temperature_flattening}, we have confirmed that the flattening also occurs in other islands and at other times.

The flattening weakens the local temperature-gradient drive of the microtearing instability and is therefore expected to promote turbulence saturation. The same saturation mechanism may thus operate with and without equilibrium magnetic shear.
% This flattening indicates that the linear drive of the microtearing modes is reduced.
% The reduced linear drive is expected to cause the energy to decay until the drive balances the dissipation, eventually leading the system to a steady state.
% Our results suggest that, even in the absence of magnetic shear, the saturation mechanism may be the same as that in microtearing turbulence with finite magnetic shear~\cite{Ajay+2023}.
%A detailed analysis of this process is left for future work.

\section{Summary}
\label{sec:summary}

We have investigated the self-organization of magnetic islands and zonal magnetic fields in collisional microtearing turbulence in the absence of magnetic shear, using gyrokinetic simulations with the \texttt{GS2} code.

\begin{figure}[!ht]
    \centering
    \includegraphics[width=0.9\linewidth]{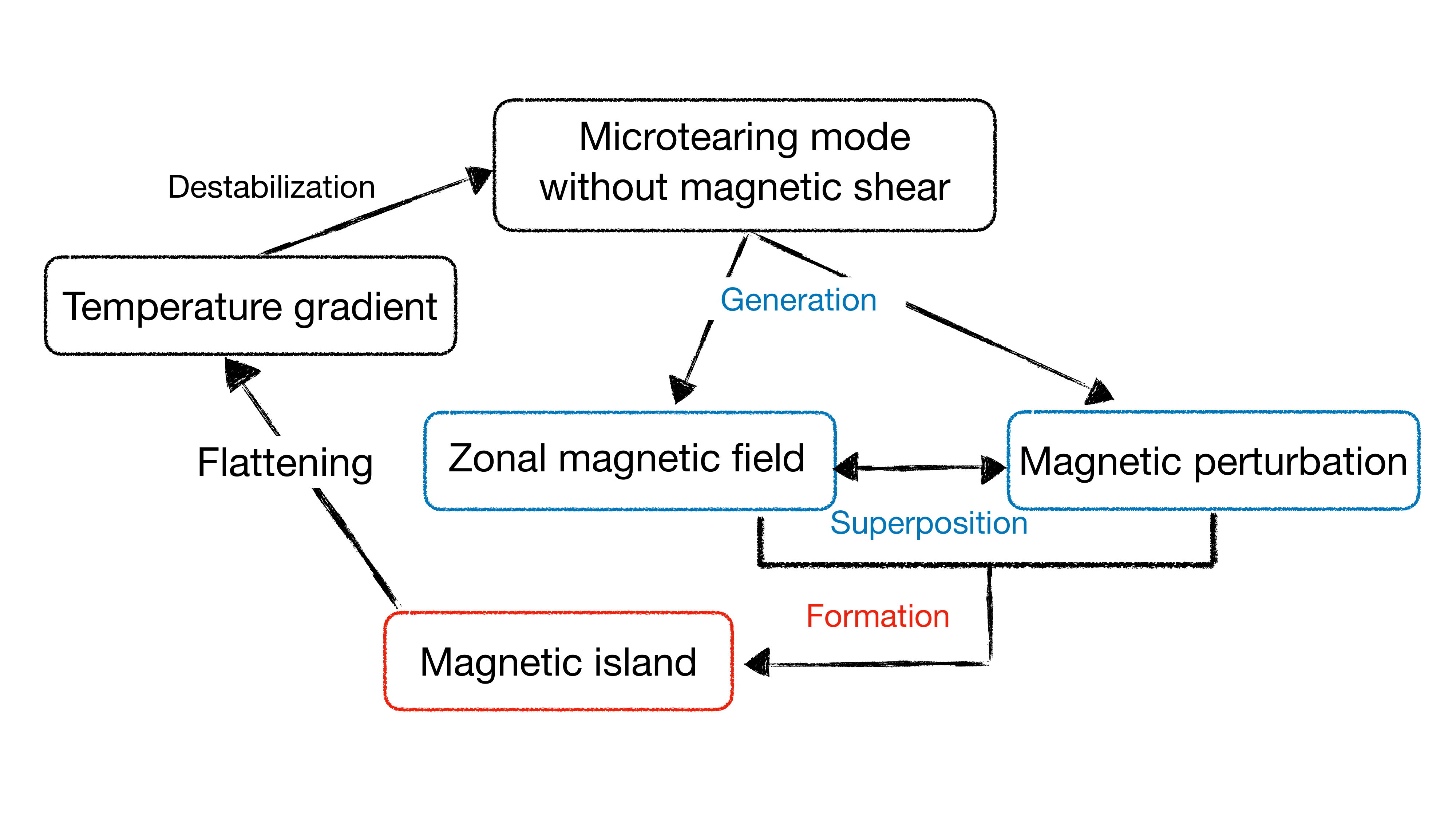}
   \caption{Schematic diagram of the self-organization process driven by microtearing modes in the absence of magnetic shear.}
    \label{fig:schematic_diagram_self-organization}
\end{figure}

%Figure~\ref{fig:schematic_diagram_self-organization} summarizes the self-organization process driven by microtearing modes in the absence of magnetic shear.
%The blue boxes represent the magnetic fields generated by the microtearing modes, while the red box represents the resulting magnetic islands.
%The microtearing modes are destabilized by the background electron temperature gradient through the time-dependent thermal force.
We have shown that the gyrokinetic free energy is dominated by the electron particle energy associated with isotropic electron temperature fluctuations.
The power balance analysis of $A_{\para}$ based on the generalized Ohm's law has demonstrated that zonal magnetic fields are generated by the nonlinear interaction between magnetic perturbations and electron temperature fluctuations.
These self-generated zonal magnetic fields act as an effective magnetic shear, allowing the magnetic perturbations to self-organize into magnetic islands even in the absence of equilibrium magnetic shear.
We have also found that, within the magnetic islands, the electron temperature fluctuations produce a gradient that nearly cancels the background temperature gradient, thereby flattening the electron temperature profile.
This flattening weakens the instability drive of the microtearing modes and is therefore expected to promote turbulence saturation, suggesting that the same saturation mechanism may operate with and without equilibrium magnetic shear.
The overall self-organization process is illustrated schematically in figure~\ref{fig:schematic_diagram_self-organization}.
A detailed investigation of the subsequent evolution in the intermediate nonlinear and saturation phases is left for future work.

The results of this paper may have implications for electromagnetic turbulence in low-magnetic-shear regions of magnetic confinement devices.
In the Wendelstein 7-X stellarator, microtearing turbulence has recently been reported in a region with nearly zero magnetic shear~\cite{Cu+2026}.
Weak or nearly zero magnetic shear is also relevant to advanced tokamak operating scenarios, such as those planned for JT-60SA~\cite{Gabriellini+2025, Wakatsuki+2023}.
In such regions, the self-organization mechanism investigated in this work may play an important role in the nonlinear evolution of microtearing turbulence.

We speculate that the role of zonal magnetic fields in the nonlinear evolution of microtearing turbulence depends on the strength of magnetic shear.
In weak-magnetic-shear regimes, self-generated zonal magnetic fields may provide an effective magnetic shear and play an important role in the nonlinear evolution, as observed in this study.
In contrast, in regimes with sufficiently strong magnetic shear, such self-organization may be suppressed, and zonal magnetic fields may play a less significant role.
A systematic investigation of the relationship between magnetic shear strength and the generation of zonal magnetic fields is therefore an important direction for future work.

%TC:ignore
\section*{Acknowledgements}
% This work was supported by JSPS KAKENHI Grant Number 22K03568.
This work was performed on “Plasma Simulator” of NIFS with the support and under the auspices of the NIFS Collaboration Research program (NIFS25KISM016).
This work was supported by a research grant from Wesco Science Promotion and the Notre Dame Seishin University research grant. 
% We would like to thank Editage [\url{http://www.editage.com}] for editing and reviewing this manuscript for English language.
%TC:endignore

\appendix
\renewcommand{\thesection}{Appendix \Alph{section}}
\counterwithin{equation}{section}
\renewcommand{\theequation}{\Alph{section}.\arabic{equation}}

\section{Gyrokinetic--Maxwell equations and fluid moments}

\subsection{Gyrokinetic--Maxwell equations}
\label{appdix_GK-M_eqn}
The non-Boltzmann part $h_{s}$ of the perturbed distribution function of species $s$ ($s = \mathrm{i}, \mathrm{e}$) obeys the gyrokinetic equation given by
\begin{align}
    \frac{\p h_s}{\p t}
    + v_{\para} \left(\frac{\bm{B}_{0}}{B_{0}}+\frac{\bm{B}_{\perp}}{B_{0}}\right)\cdot \nabla h_{s}
    +\bm{u}_{E}\cdot \nabla_{\perp} h_{s}
    = \frac{q_{s} f_{0s}}{T_{0s}}\left(\frac{\p}{\p t} + \bm{v}_{\ast,s}^{T}(v) \cdot \nabla_{\perp} \right) \Rblacket{\chi}{s} + C(h_{s}),
\label{eq:gyrokinetic_eqn}
\end{align}
where $C(h_{s})$ is the collision term. 
The $\bm{E}\times\bm{B}$ drift velocity is defined by $\bm{u}_{E} = (\bm{B}_{0}/B_{0})\times \nabla_{\perp}\Rblacket{\phi}{s}/B_{0}$. $\bm{B}_{\perp} = \nabla_{\perp} \Rblacket{A_{\para}}{s} \times (\bm{B}_0/B_{0})$ is the magnetic perturbation associated with the parallel magnetic vector potential $A_{\para}$.
The gyrokinetic potential is defined as $\chi = \phi - v_{\para} A_{\para}$. Here, $\delta B_{\para}$ is neglected.
The angle brackets $\Rblacket{\cdot}{s}$ denote the gyroaverage at fixed guiding-center position $\bm{R}_{s} = \bm{r} + \bm{v}\times(\bm{B}_{0}/B_{0})/\Omega_{\mathrm{c}s}$, where $\bm{r}$ is the particle position and $\Omega_{\mathrm{c}s} = q_{s} B_{0}/m_{s}$ is the cyclotron frequency.
The equilibrium distribution function $f_{0s}$ is the local Maxwellian,
\begin{align}
    f_{0s} = \frac{n_{0s}}{\pi^{3/2} v_{\mathrm{th},s}^{3}} \exp\left(-\frac{v^{2}}{v_{\mathrm{th},s}^{2}}\right),
\end{align}
with $v_{\mathrm{th},s} = \sqrt{2T_{0s}/m_{s}}$ and $v = |\bm{v}|$.
The velocity-dependent diamagnetic drift velocity is given by
\begin{align}
    \bm{v}_{\ast,s}^{T}(v) = -\frac{T_{0s}}{q_{s}B_{0}}\left[\frac{1}{L_{n_{0s}}} + \left(\frac{v^{2}}{v_{\mathrm{th},s}^{2}} - \frac{3}{2}\right)\frac{1}{L_{T_{0s}}}\right]\hat{\bm{y}},
\end{align}
where the density and temperature gradients are characterized by $L_{n_{0s}}^{-1} = - \p_{x} \ln n_{0s}$ and $L_{T_{0s}}^{-1} = - \p_{x} \ln T_{0s}$, respectively.
These gradients give rise to the diamagnetic drift frequencies, defined as
\begin{align}
    \omega_{\ast,n_{0s},T_{0s}} \equiv 
    -\frac{|q_{s}|}{q_{s}}\frac{k_{y} \rho_{s}}{2} \frac{v_{\mathrm{th},s}}{L_{n_{0s},T_{0s}}},
\end{align}
where $\rho_{s}=v_{\mathrm{th},s}/|\Omega_{\mathrm{c}s}|$ is the Larmor radius and $k_{y}$ is the wavenumber in the $y$ direction.
Since the pressure gradient is given by $L_{P_{0s}}^{-1}=L_{n_{0s}}^{-1} + L_{T_{0s}}^{-1}$, the diamagnetic drift frequency associated with $L_{P_{0s}}^{-1}$ is defined as $\omega_{\ast,s} \equiv \omega_{\ast,n_{0s}} + \omega_{\ast, T_{0s}}$.

The gyrokinetic equation~\eqref{eq:gyrokinetic_eqn} is closed by the quasi-neutrality condition
\begin{align}
    \sum_{s} \frac{q_{s}^{2} n_{0s}}{T_{0s}}\,\phi
    = \sum_{s} q_{s} \int \langle h_{s} \rangle_{\bm{r}}\,\dv,
\label{eq:quasineutrality}
\end{align}
and the parallel component of Amp\`ere's law
\begin{align}
    -\nabla_{\perp}^{2} A_{\para}
    = \mu_{0} \sum_{s} q_{s} \int v_{\para} \langle h_{s} \rangle_{\bm{r}}\,\dv,
\label{eq:ampere}
\end{align}
which determine $\phi$ and $A_{\para}$, respectively. 
Here, $\mu_{0}$ is the vacuum permeability, and the brackets $\langle \cdot \rangle_{\bm{r}}$ denote the gyroaverage at fixed particle position $\bm{r}$.

\subsection{Fluid moments}
\label{appdix:fluidmoms}
Some low-order velocity moments of $\delta f_{s}$ are given by
\begin{align}
    \tilde{n}_{s} &= \int \delta f_{s}\, \mathrm{d}\bm{v}, \label{eq:moment_n}\\
    n_{0s} \tilde{u}_{\parallel,s} &= \int v_{\parallel}\, \delta f_{s}\, \mathrm{d}\bm{v}, \label{eq:moment_u}\\
    \tilde{p}_{\parallel,s} &= \int m_{s} v_{\parallel}^{2}\, \delta f_{s}\, \mathrm{d}\bm{v}, \label{eq:moment_ppara}\\
    \tilde{p}_{\perp,s} &= \int \frac{m_{s} v_{\perp}^{2}}{2}\, \delta f_{s}\, \mathrm{d}\bm{v}, \label{eq:moment_pperp}
\end{align}
where $\tilde{n}_{s}$, $\tilde{u}_{\parallel,s}$, $\tilde{p}_{\parallel,s}$, and $\tilde{p}_{\perp,s}$ are the density, parallel flow velocity, and parallel and perpendicular pressure fluctuations, respectively.
To first order in the fluctuations, the pressure fluctuations are decomposed as $\tilde{p}_{\parallel,s} = \tilde{n}_{s} T_{0s} + n_{0s} \tilde{T}_{\parallel,s}$ and $\tilde{p}_{\perp,s} = \tilde{n}_{s} T_{0s} + n_{0s} \tilde{T}_{\perp,s}$.
The parallel and perpendicular temperature fluctuations are therefore given by
\begin{align}
    \tilde{T}_{\parallel,s} &= \frac{1}{n_{0s}} \int \left( m_{s} v_{\parallel}^{2} - T_{0s} \right) \delta f_{s}\, \mathrm{d}\bm{v}, 
    \label{eq:moment_Tpara}\\
    \tilde{T}_{\perp,s} &= \frac{1}{n_{0s}} \int \left( \frac{m_{s} v_{\perp}^{2}}{2} - T_{0s} \right) \delta f_{s}\, \mathrm{d}\bm{v}. 
    \label{eq:moment_Tperp}
\end{align}
The isotropic temperature fluctuation is defined as
\begin{equation}
    \tilde{T}_{s} = \frac{\tilde{T}_{\parallel,s} + 2\tilde{T}_{\perp,s}}{3}
    = \frac{1}{n_{0s}} \int \left( \frac{m_{s} v^{2}}{3} - T_{0s} \right) \delta f_{s}\, \mathrm{d}\bm{v}.
    \label{eq:moment_T}
\end{equation}
%which corresponds to the linearized isotropic pressure $\tilde{p}_{s} = (\tilde{p}_{\parallel,s} + 2\tilde{p}_{\perp,s})/3 = \tilde{n}_{s} T_{0s} + n_{0s} \tilde{T}_{s}$.

\section{Power balance equation for $A_{\para}$}
\label{appdix_ohms_law}
In this appendix, we derive the power balance equation for $A_{\para}$ used in section~\ref{subsec:magnetic_island}.
We consider the electron gyrokinetic equation~\eqref{eq:gyrokinetic_eqn} in the drift-kinetic approximation and in the absence of magnetic shear. Taking its $v_{\para}$ moment yields the electron parallel momentum equation.
Solving the momentum equation for $\p A_{\para}/\p t$, we obtain the generalized Ohm's law,
\begin{align}
    \frac{\p A_{\para}}{\p t}
    =
    -\frac{m_{\ele}}{q_{\ele}}\frac{\diff \tilde{u}_{\para,\ele}}{\diff t}
    -\bm{u}_{E}\cdot \nabla_{\perp} A_{\para}
    &-\frac{T_{0\ele}}{q_{\ele}}\frac{\bm{B}_{\perp}}{B_{0}} \cdot \nabla_{\perp}\left(\frac{\tilde{n}_{\ele}}{n_{0\ele}}+\frac{\tilde{T}_{\para,\ele}}{T_{0\ele}}\right) \nonumber \\
    &-v_{\ast,n_{0\ele}}(1+\eta_{\ele})\frac{\p A_{\para}}{\p y}
    - E_{\para}^{\mathrm{coll}},
    \label{eq:parallel_electric_field}
\end{align}
where $\diff /\diff t = \p/\p t + \bm{u}_{E}\cdot \nabla_{\perp}$ is the convective time derivative, $v_{\ast,n_{0\ele}} = \omega_{\ast,n_{0\ele}}/k_{y}$ is the electron diamagnetic drift velocity, and $E_{\para}^{\mathrm{coll}}$ represents the contribution of the collision term to the parallel electric field.
\eqref{eq:parallel_electric_field} has the dimensions of the parallel electric field, as follows from the parallel component of Faraday's law, $\p A_{\para}/\p t = -E_{\parallel}$.
The temperature gradient normalized by the density gradient is denoted by $\eta_{\ele}\equiv L_{n_{0\ele}}/L_{T_{0\ele}}$.
Since the anisotropic temperature fluctuation is negligible compared with the isotropic one (figure~\ref{fig:total_energy_evolution}), we take $\tilde{T}_{\para,\ele}=\tilde{T}_{\ele}$.

We expand \eqref{eq:parallel_electric_field} in a Fourier series in the $y$ direction.
Multiplying the $k_{y}$ component by the complex conjugate $A^{\ast}_{\para,k_{y}}(x,t)=A_{\para}^{\ast}(x,k_{y},t)$ and taking the real part, we obtain the power balance equation for $A_{\para}$:
\begin{align}
    \frac{1}{2} \frac{\p}{\p t} \left|A_{\para,k_{y}} (x,t)\right|^2
    =
    \sum_{j} \Re \left[A_{\para,k_{y}}^{\ast}(x,t) \, N_{k_{y}}^{j}(x,t) \right],
    \label{eq:power_eqn_apar}
\end{align}
where $j$ labels the following terms:
\begin{align}
    N^{\tilde{u}_{\para,\ele}}_{k_{y}}
    &= -\frac{m_{\ele}}{q_{\ele}}
        \left(\frac{\diff \tilde{u}_{\para,\ele}}{\diff t}\right)_{k_{y}}, \\
    N^{\phi,A_{\para}}_{k_{y}}
    &= -\left(\bm{u}_{E}\cdot\nabla_{\perp}A_{\para}\right)_{k_{y}},  \\
    N^{A_{\para},\tilde{n}}_{k_{y}}
    &= -\frac{T_{0\ele}}{q_{\ele}}
        \left(\frac{\bm{B}_{\perp}}{B_{0}}\cdot\nabla_{\perp}\frac{\tilde{n}_{\ele}}{n_{0\ele}}\right)_{k_{y}}, \\
    N^{A_{\para},\tilde{T}}_{k_{y}}
    &= -\frac{T_{0\ele}}{q_{\ele}}
        \left(\frac{\bm{B}_{\perp}}{B_{0}}\cdot\nabla_{\perp}\frac{\tilde{T}_{\ele}}{T_{0\ele}}\right)_{k_{y}}, \\
    N^{\mathrm{coll}}_{k_{y}}
    &= - E_{k_{y}}^{\mathrm{coll}}(x,t).
\end{align}
$(\cdot)_{k_{y}}$ denotes the Fourier component at $k_{y}$, which includes the full nonlinear mode coupling.
The linear diamagnetic drift term in \eqref{eq:parallel_electric_field} does not contribute to \eqref{eq:power_eqn_apar}.
This term is proportional to $A_{\para}\p A_{\para}/\p y = \p(A_{\para}^{2}/2)/\p y$. The total $y$-derivative vanishes upon averaging over $y$, and the diamagnetic drift term only represents the propagation of $A_{\para}$ in the $y$ direction.
%The linear diamagnetic drift term in \eqref{eq:parallel_electric_field} does not appear in \eqref{eq:power_eqn_apar}, because its Fourier component, $-\mathrm{i}k_{y}v_{\ast,n_{0\ele}}(1+\eta_{\ele})A_{\para}$, gives a purely imaginary product with $A_{\para}^{\ast}$.

Averaging \eqref{eq:power_eqn_apar} over $x$, we obtain
\begin{align}
    \frac{1}{2}\frac{\p P_{A_{\para,{k_{y}}}}}{\p t}
    =
    \sum_{j} F_{k_{y}}^{j},
    \label{eq:power_eqn_apar_xavg}
\end{align}
with
\begin{align}
    &P_{A_{\para,{k_{y}}}} \equiv \left\langle \left|A_{\para,k_{y}}(x,t)\right|^{2} \right\rangle_{x}, \\
    &F_{k_{y}}^{j} \equiv \left\langle \Re \left[A^{\ast}_{\para,k_{y}}(x,t)\, N_{k_{y}}^{j}(x,t)\right]\right\rangle_{x},
\end{align}
where $\langle\cdot\rangle_{x}$ denotes the $x$-average.
Dividing \eqref{eq:power_eqn_apar_xavg} by $P_{A_{\para,{k_{y}}}}$ and introducing the growth rate $\gamma_{k_{y}} \equiv \p \ln P_{A_{\para,{k_{y}}}}^{1/2}/\p t$, we obtain
\begin{align}
    \gamma_{k_{y}} = \sum_{j} \gamma_{k_{y}}^{j},
    \label{eq:growth_rate_decomposition}
\end{align}
where $\gamma_{k_{y}}^{j} \equiv F_{k_{y}}^{j}/P_{A_{\para,{k_{y}}}}$ is the contribution of each term to the growth rate.

\section*{References}
\bibliography{bib/references}

\bibliographystyle{iopart}  %iopart, jphysicsB
\end{document}